\documentclass{desyproc}

\usepackage{wrapfig}
\usepackage{sidecap}
\usepackage{subfigure}

\begin{document}

%------------------------------------
\title{\vspace{-3cm}{\small \hfill{MPP-2011-124}}\\[1.4cm]
Cosmological constraints on thermal relic
\\axions and axion-like particles}

%for single authors the superscripts are optional
\author{{\slshape Davide Cadamuro$^1$, Javier Redondo$^1$}\\[1ex]
$^1$Max-Planck-Institut f\"ur Physik, Munich, Germany}

% if the proceedings are available online (e.g. at Indico)
% please enter the contribution ID or file_name below for the DOI
%\contribID{32}
\contribID{familyname\_firstname}

% TO THE CONFERENCE EDITORS: 
% please update the following information      
% before sending the template to the authors
% \confID{800}  % if the conference is on Indico uncomment this line
\desyproc{DESY-PROC-2011-04}
\acronym{Patras 2011} % if you want the Acronym in the page footer uncomment this line
\doi  % if there is an online version we will register DOIs

\maketitle

\begin{abstract}
Cosmological precision data can be used to set very strict constraints on Axions and Axion-like particles (ALPs) produced thermally in the big bang. 
We briefly review the known bounds and propose two new constraints for Axions and ALPs decaying in the early universe, based upon the concomitant dilution of baryon and neutrino densities, using WMAP7 and other cosmological data.
\end{abstract}

\section{Cosmology of decaying axions and ALPs}
\label{sec:cosmology}

The axion is a pseudo-Goldstone boson arising from the spontaneous breaking of the global axial Peccei-Quinn $U(1)$ symmetry 
proposed to solve the strong CP problem~\cite{Kim:1986ax}.
The effective lagrangian relevant for this work includes only their coupling to photons      
%%%%%
\begin{equation}\label{eq:axionlagrangian}
\mathcal{L}_{\rm eff}^{a}=\frac{1}{2}\partial^\mu a \partial_\mu a - \frac{1}{2}m_a^2 a^2 -a\frac{g_{a\gamma}}{4}F^{\mu\nu}\widetilde{F}_{\mu\nu} . 
\end{equation}
Considering couplings to electrons and other leptons changes only slightly our conclusions. 
The mass and the photon coupling are calculable in chiral perturbation theory as~\cite{Kim:1986ax},
%%%%%
\begin{eqnarray}
m_a & = & \frac{f_\pi m_\pi}{f_a}\frac{\sqrt{m_u m_d}}{m_u+m_d}\simeq 6\,{\rm eV}\left(\frac{10^6~\rm GeV}{f_a}\right)\\
g_{a\gamma}&=&\frac{\alpha}{2\pi f_a}\left(\frac{E}{N}-\frac{2}{3}\frac{m_u+4 m_d}{m_d+m_u}\right)\equiv \frac{1.9\,\alpha\, \delta}{2\pi f_a}.
\end{eqnarray}
and are related to each other via $\delta$, a free ${\cal O}(1)$ parameter. 
The two-photon coupling allows the decay $a\rightarrow\gamma\gamma$, at the rate
\begin{equation}
\label{decayrate}
\Gamma_{a\gamma\gamma}= \frac{g_{a\gamma}^2 m_a^3}{64 \pi} . 
\end{equation} 

The two photon coupling allows to axions thermalize in the early universe via the Primakoff process $q^\pm+\gamma\leftrightarrow q^\pm+a$. However, at some temperature axions decouple. Additional axion interactions with pions or electrons can lower the decoupling temperature, but are not relevant for this discussion. This thermal axion population behaves as hot dark matter, which is currently disfavoured, implying the constraint $m_a<0.7$ eV~\cite{Hannestad:2010yi}.

The axion lifetime $\tau=\Gamma_{a\gamma\gamma}^{-1}\sim 10^{24}~{\rm s} \,
({\rm eV}/m_a)^5\delta^{-2}$ becomes shorter than the age of the universe if the mass exceeds $\sim20$ eV and the above bound looses his meaning.  
We have therefore explored the possibility that, above a certain mass, the axion could disappear from the universe without leaving any observable trace~\cite{Cadamuro:2010cz}. 
Still, the decay of axions would affect BBN, the neutrino temperature and the CMB. 
The exquisite precision of cosmological data sets permits to constrain the axion parameters. 
All in all, we found that axions are allowed by cosmology to have $m_a>300$ keV~\cite{Cadamuro:2010cz}.

These cosmological limits become even more interesting in the case of axion-like particles (ALPs): pseudo-scalars having similar properties of the axion except the relation between the mass and the photon coupling, which are now free to span the whole parameter space~\cite{Masso:1995tw}. In a recent paper~\cite{Cadamuro:2011fd}, we have extended to ALPs the axion cosmological bounds we found in~\cite{Cadamuro:2010cz}. 
The summary of the constraints we found is in Fig.~\ref{fig:1}, together with some other relevant cosmological and astrophysical bounds.

The most interesting bounds arise because of the entropy released during the decay. 
As the decay products are photons (and in some axion models also electrons), {\em this entropy is shared 
only among the species in thermal equilibrium with the electromagnetic radiation}.
If axions or ALPs decay when inverse reactions are efficient, $\gamma\gamma\to a$, they do it in local thermal equilibrium (LTE). 
This is the case if ALPs decay when the temperature exceeds greatly their mass. 
The raise in entropy follows from entropy conservation by counting the relevant 
 relativistic degrees of freedom $g_{*S}$ before and after the disappearance, like one usually does for the $e^\pm$ annihilation. 
For instance, if axions or ALPs decay affects only photons, the photon entropy is increased by a factor $(2+1)/2=3/2$. 
   
The other extreme possibility is that the temperature is smaller than the mass when axions or ALPs decay, which implies far out-of-equilibrium decay. In this case the energy density in scalars can even dominate the universe expansion thrusting in a huge amount of entropy. In this case, one finds~\cite{Kolb:1990vq}
\begin{equation}
\label{OUT}
\frac{S_{f}}{S_{i}}\sim\frac{ m_a R}{\sqrt{m_{\rm Pl}\Gamma_{a\gamma\gamma}} }
\end{equation}
where $R$ is the number of ALPs per photon before the decay. 
Solving the Boltzmann equations we exactly calculated the entropy emitted during the decay and how this influences the thermodynamics of the other species.

\section{Number of effective neutrinos}
\label{sec:neutrinos}

\begin{figure}[tb]
\centering
\begin{minipage}[t]{0.47\linewidth}
%\centering
\flushleft
\includegraphics[width=6.5cm]{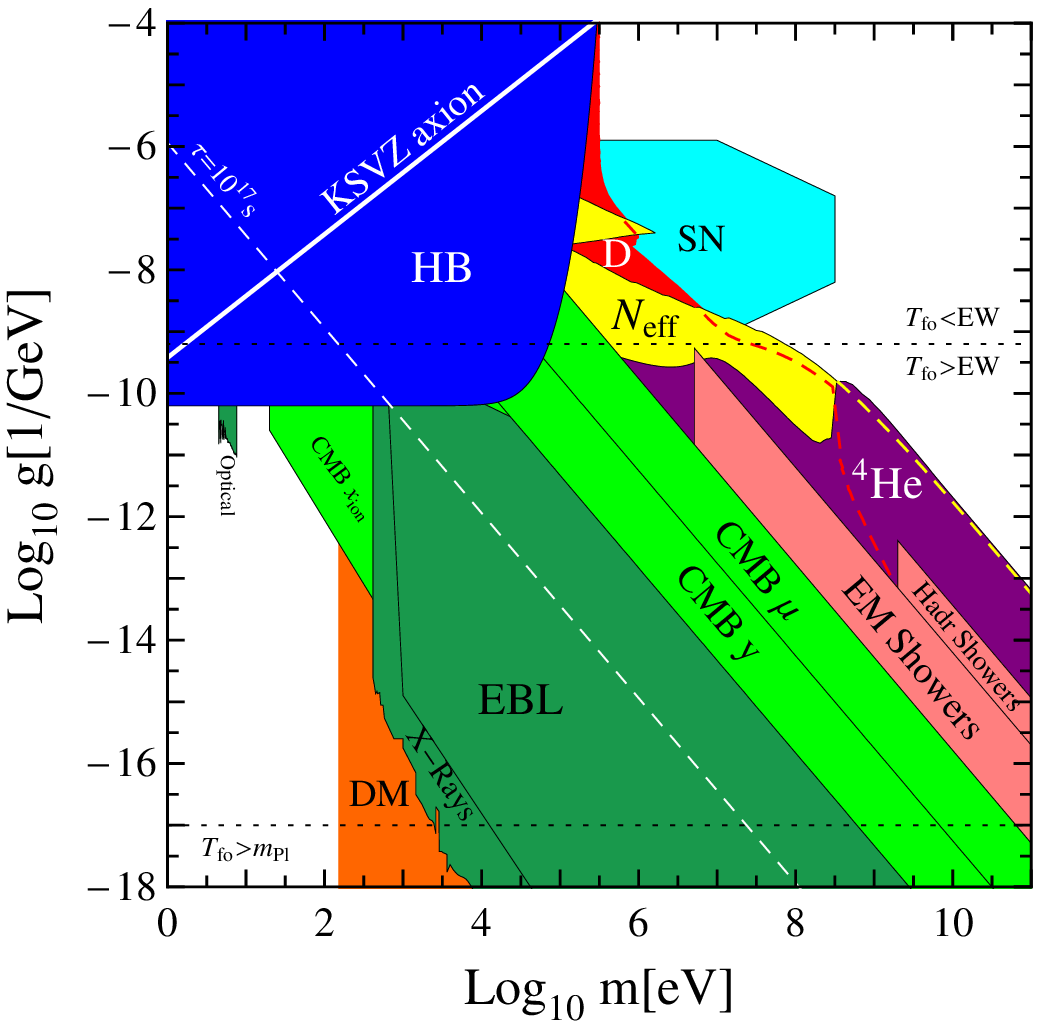}
\vspace{-.1cm}
  { \caption{\small Summary of ALP bounds in the mass-coupling plane.  The ALP relic density exceeds the observed value in the region labelled DM;  EBL, Optical and X-Rays regions are excluded by the non observation of photons from relic ALP decay; The black-body spectrum of CMB would be unacceptably distorted by ALP decays in the regions labelled CMB.
$N_{\rm eff}$, D, Showers and He are described in the text; 
SN and HB are astrophysical bounds. See~\cite{Cadamuro:2011fd} for details.  }}
   \label{fig:1}
\end{minipage}
%%%%%%%%%%%%%%%%%%%%%%%%%%%%%%%%%%
\hspace{0.4cm}
%%%%%%%%%%%%%%%%%%%%%%%%%%%%%
\begin{minipage}[t]{0.47\linewidth}
%\centering
\flushleft
\includegraphics[width=6.5cm]{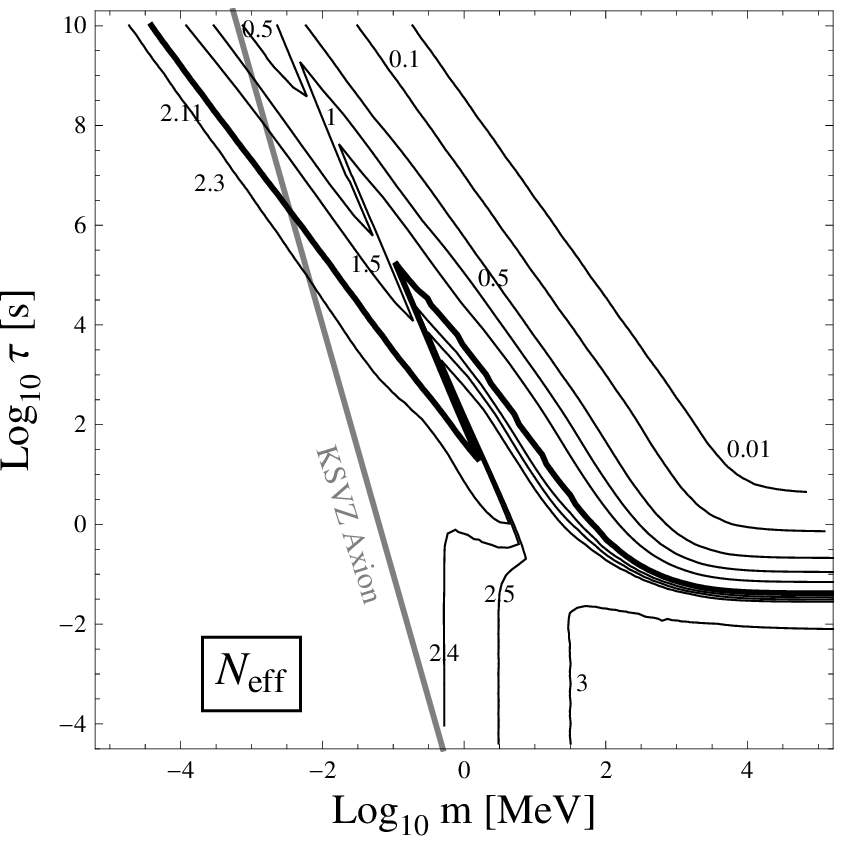}
\vspace{-.1cm}
   {\caption{ \small Isocontours of $N_{\rm eff}$, as a function of the ALP mass $m$ and the lifetime $\tau$. 
The excluded region lies above the thick black line.
LTE decay (lower-left corner) produces no bound for $N_{\rm eff}$. 
In the upper-right corner the decay is out-of-equilibrium and during ALP domination: $\nu$s are almost washed out. 
Decreasing $\tau$ the bound becomes less severe: it completely disappears $\tau$ shorter than the $\nu$ freeze-out~\cite{Cadamuro:2011fd}. 
For the KSVZ axion $m_a\gtrsim 3$ keV is allowed at 99\% C.L..~\cite{Cadamuro:2010cz}. }}
   \label{fig:2}
\end{minipage}
\end{figure}

If axions or ALPs decay after the decoupling of neutrinos, the latter can not be heated by the radiation released and the final ratio $T_{\nu}/T_{\gamma}$ would be lower than the standard value $(4/11)^{1/3}$. 
The number of effective neutrinos is defined as 
\begin{equation}
N_{\rm eff} = \frac{8}{7}
{\left(\frac{11}{4}\right)}^{4/3}\frac{\rho_\nu}{\rho_\gamma}
\end{equation}
to measure the neutrino energy density.
From WMAP7, the 7th release of the SSDS and the measurement of $H_0$ by the HST, we obtained the 
following limits on $N_{\rm eff}$~\cite{Cadamuro:2010cz}
\begin{equation}\label{eq:neffconstraint}
N_{\rm eff}>
\begin{cases}
{2.70}&\hbox{at 68\% C.L.}\\
{2.39}&\hbox{at 95\% C.L.}\\
{2.11}&\hbox{at 99\% C.L.}\ ,
\end{cases}
\end{equation}
assuming a prior $N_{\rm eff}<3$ because we assume only $3$ standard neutrinos and in the ALP decay scenario sketched the ALP decay can only increase $\rho_\gamma$, i.e. lower $N_{\rm eff}$.
We plot our results in Fig.~\ref{fig:2} in the ALP mass and lifetime plane. Axion models with $\delta=1$ lie along the steep line labelled KSVZ.

%%%%%%%%%%%%%%%%%%%%%%%%%%%%%
%%%%%%%%%%%%%%%%%%%%%%%%%%%%

%%%%%%%%%%%%%%%%%%%%%%%%%%%%%%%%%%
\begin{figure}[tb]
\centering
\subfigure[Deuterium]{\label{fig:3}\includegraphics[width=6.5cm]{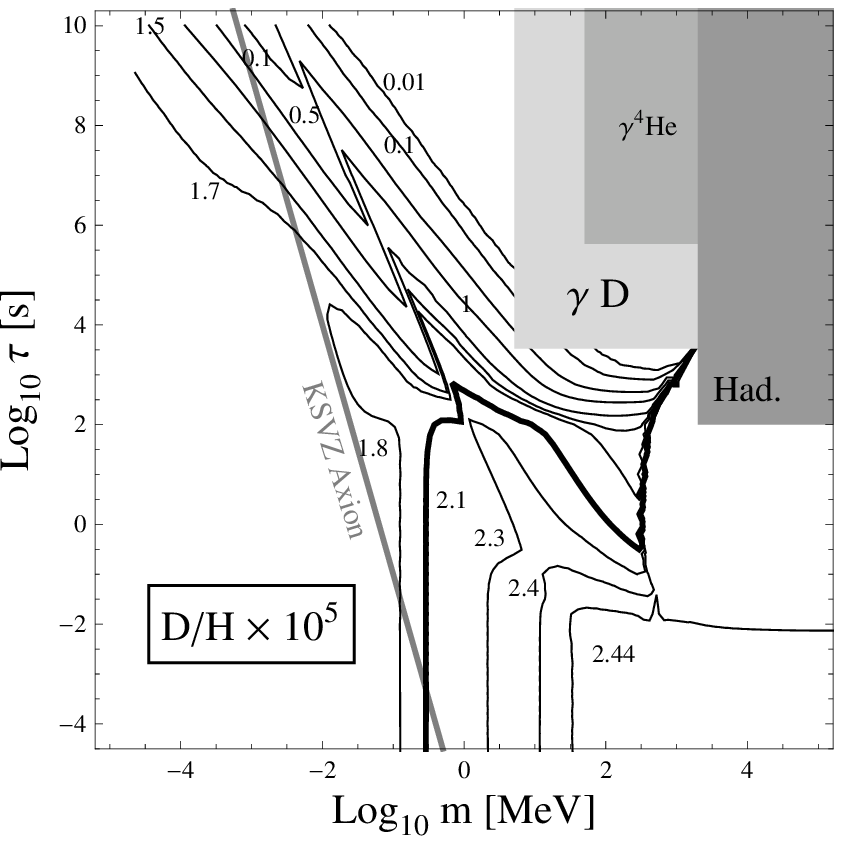}}
\hspace{0.4cm}
\subfigure[Helium]{\label{fig:4}\includegraphics[width=6.5cm]{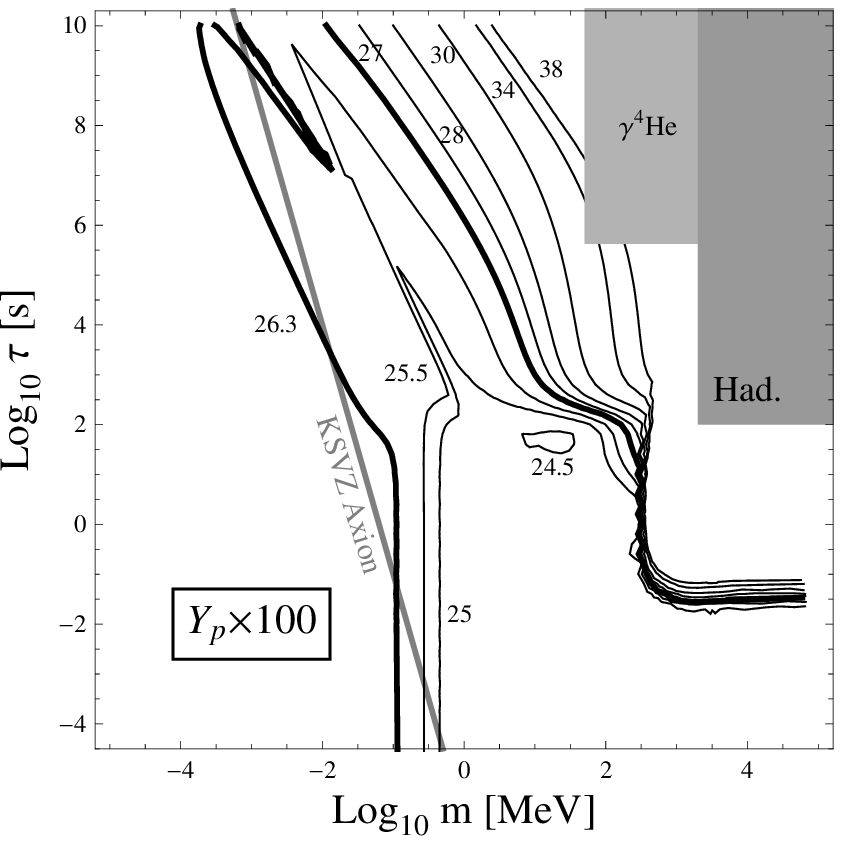}}
\vspace{-.2cm}
\caption{\small Deuterium to proton abundances ratio (left) and helium mass fraction (right) from BBN in the decaying ALP cosmology, as in Fig.~\ref{fig:2}.
   The LTE decay (lower-left corner) case is enough to exclude masses below 300 keV at 99\% C.L.~because of D underproduction, also for the KSVZ axion~\cite{Cadamuro:2010cz}. 
   Massive ALPs dissociate nuclei through EM and hadronic cascades and are excluded in the grey regions. 
   BBN happens in a low $N_{\rm eff}$ universe for $\tau<100$ s, which generally translates into helium overproduction. 
   For shorter lifetimes ($\tau\sim0.1-100$ s) the decay produces charged pions that enhance the neutron-to-proton ratio at BBN, resulting in a high D/H and He mass fraction~\cite{Cadamuro:2011fd}.}
\label{fig:BBN}
\end{figure}
%%%%%%%%%%%%%%%%%%%%%%%%%%%%%%%%%%

\section{Primordial Nucleosynthesis}
\label{sec:BBN}

The photons emitted in ALP decays decrease the baryon to photon ratio $\eta_B$, which is well measured from CMB to be $\eta_B^{\rm CMB}=6.23\times10^{-10}$. 
This value agrees with what inferred from standard BBN. 
If a relic population decays before the CMB release but after BBN it means BBN happened with a higher $\eta_B$, i.e. $\sim(S_f/S_i) \eta_B^{\rm CMB}$. 
The LTE value $S_f/S_i=3/2$ is enough to create a disagreement between BBN simulations and the observation of primordial elements. 
ALPs with longer lifetimes decay increasingly out-of-equilibrium and rise the tension,  
unless a decrease in $R$ compensates the growth of $(\Gamma_{a\gamma\gamma})^{-1/2}$, c.f.~Eq.~\eqref{OUT} (the sharp features of our results correspond to the large decrease in $R$ when ALPs decouple around the QCD phase transition).  
Deuterium is the most affected nucleus in a high $\eta_B$ BBN, where it can be more efficiently processed into heavier elements: it is the best observable to test $S_f/S_i$. 
Values of $D/H< 2.2\times10^{-5}$ are presently excluded at 99\% C.L., which allows only axions with mass larger than $300$ keV~\cite{Cadamuro:2010cz}. 
The results for the ALP case are plotted in Figs.~\ref{fig:BBN}. 
Higher mass ALPs initiate EM and hadronic showers, which dissociate nuclei. 
Also ALPs decays produce charged pions, whose presence alters the neutron/proton equilibrium. In general, they enhance $n_n/n_p$, which can even reach values $\sim$ 1. 
This translates into high He and D/H yields~\cite{Cadamuro:2011fd}.
The $N_{\rm eff}$, D and He limits are presently the most constraining ones for shorter ALP lifetimes.

\bibliographystyle{plain}

\end{document}